# Out of sight out of mind: Perceived physical distance between the observer and someone in pain shapes observer's neural empathic reactions

Arianna Schiano Lomoriello[1], Federica Meconi[2], Irene Rinaldi[1], & Paola Sessa[1, 3] *

[1] Department of Developmental and Social Psychology, University of Padova, Via Venezia 8, 35121, Padova, Italy.

[2] School of Psychology, University of Birmingham, Birmingham, United Kingdom.

[3] Padova Neuroscience Center (PNC), University of Padova, Padova, Italy.

**\* Corresponding author:** Paola Sessa, Department of Developmental and Social Psychology, University of Padova, Via Venezia 8, 35121, Padua, Italy.

paola.sessa@unipd.it

**Abstract**

Social and affective relations may shape empathy to others' affective states. Previous studies also revealed that people tend to form very different mental representations of stimuli on the basis of their physical distance. In this regard, embodied cognition and embodied simulation propose that different physical distances between individuals activate different interpersonal processing modes, such that close physical distance tends to activate the interpersonal processing mode typical of socially and affectively close relationships. In Experiment 1, two groups of participants were administered a pain decision task involving upright and inverted face stimuli painfully or neutrally stimulated, and we monitored their neural empathic reactions by means of event-related potentials (ERPs) technique. Crucially, participants were presented with face stimuli of one of two possible sizes in order to manipulate retinal size and perceived physical distance, roughly corresponding to the close and far portions of social distance. ERPs modulations compatible with an empathic reaction were observed only for the group exposed to face stimuli appearing to be at a close social distance from the participants. This reaction was absent in the group exposed to smaller stimuli corresponding to face stimuli observed from a far social distance. In Experiment 2, one different group of participants was engaged in a match-to-sample task involving the two-size upright face stimuli of Experiment 1 to test whether the modulation of neural empathic reaction observed in Experiment 1 could be ascribable to differences in the ability to identify faces of the two different sizes. Results suggested that face stimuli of the two sizes could be equally identifiable. In line with the Construal Level and Embodied Simulation theoretical frameworks, we conclude that perceived physical distance may shape empathy as well as social and affective distance.

1. **Introduction**

Humans are endowed with an extraordinary ability to share and understand the affective states of others, and this is vital as it allows appropriate social interactions and relationships with others. This ability, known as *empathy*, is multifaceted since consisting of several aspects, including emotion contagion, empathic accuracy, concern for others, self-other distinction, emotion regulation and perspective taking (Decety and Jackson, 2004, 2006; Preston and de Waal, 2002; Zaki and Ochsner, 2012).

The present investigation aimed at exploring whether the physical distance between an observer and an individual in a particular affective state (induced by a painful stimulation) is a critical factor in modulating the magnitude of an empathic neural reaction in the observer.

In the field of social and affective neuroscience, investigation has indeed mostly focused on empathy toward others' pain (Astolfi et al., 2005; Decety and Lamm, 2007; Fan and Han, 2008; Li and Han, 2010; Meconi, Vaes, and Sessa, 2015; Sessa and Meconi, 2015; Sessa, Meconi, and Han, 2014; Sheng and Han, 2012; Sheng, Han, and Han, 2016; Singer et al., 2006). In this context, most of the proposed theoretical frameworks have conceived empathy as comprised of at least two components, widely independent and dissociable, both functionally and anatomically (Decety and Lamm, 2007; Sessa, Meconi, Castelli, and Dell'Acqua, 2014; Zaki and Ochsner, 2012). One of the components is termed affective empathy or experience sharing − mainly based on neural resonance mechanisms − and the other component is termed cognitive empathy − mainly based on mental state attribution ability (Decety and Lamm, 2007; Zaki, 2013; Zaki and Ochsner, 2012). Notably, this functional distinction corresponds to an anatomical dissociation such that affective empathy has its neural substrate in regions previously associated with the mirror neuron system (premotor cortex and inferior parietal lobule) and with the limbic system (anterior cingulate cortex and anterior insula), while the neural underpinnings of the cognitive component of empathy, related to mentalizing, are in regions associated with the Theory of Mind, including medial prefrontal cortex, temporal poles, precuneus and temporo-parietal junction (see Zaki and Ochsner, 2012; see also, e.g., Amodio and Frith, 2006; Decety, 2011; Fan, Duncan, de Greck, and Northoff, 2011; Lamm and Singer, 2010; Rizzolatti and Sinigaglia, 2010; Saxe and Kanwisher, 2003; Shamay-Tsoory, Aharon-Peretz, and Perry, 2009).

A notable aspect of the human ability to experience empathy toward other people's affective states and emotions is that it may be shaped by a variety of factors, including the characteristics of the observer and those of the individual experiencing a particular affective condition (Blair, 2005; Dapretto et al., 2006; Davis, 1983; Harris and Fiske, 2006; Hein, Silani, Preuschoff, Batson, and Singer, 2010; Philip et al., 2012; Wagner, Kelley, and Heatherton, 2011) or also the affective and social relationship existing between the observer and the other individual, such that at least part of the brain network underlying empathy (i.e., anterior cingulate cortex and anterior insula) is strongly activated in those cases when the partner, rather than a stranger, is experiencing a pain stimulation (Singer et al., 2004), or in cases when the other is an individual with whom the observer has established a relationship of trust rather than distrust (Singer et al., 2006), or when the individual experiencing pain belongs to the observer's ethnic group rather than a different ethnic group (Avenanti, Sirigu, and Aglioti, 2010; Contreras-Huerta, Baker, Reynolds, Batalha, and Cunnington, 2013; Contreras-Huerta, Hielscher, Sherwell, Rens, and Cunnington, 2014; Sessa, et al., 2014; Xu, Zuo, Wang, and Han, 2009). In brief, the chance that an empathic reaction will be triggered and its magnitude depend on the nature of the social and affective relationships that binds people.

Interestingly, social and affective relationships are often designated in terms of "distance", and just as for the physical distance, the terms "close" and "distant" tend to be used in the context of relationships, for example, associating them with an intimate friend or with a relative almost unknown to us, respectively (Lakoff and Johnson, 1980; Lakoff and Mark, 1999). In this vein, it is possible to conceive social and affective relationships between individuals as if they were mapped onto a sort of virtual space. Support in favor of this proposal comes, for instance, from a functional magnetic resonance imaging (i.e., fMRI) study by Yamakawa, Kanai, Matsumura, and Naito (2009) who asked their participants, in two different tasks, to evaluate social compatibility with presented individuals' faces and to evaluate physical distance of inanimate objects. The rationale for the implementation of these two tasks was that if evaluation of both psychological and physical distances has a common functional and neural substrate, one would expect to observe an overlapping activation in those brain regions involved in the representation of the egocentric physical space (Naito et al., 2008; Neggers, Van der Lubbe, Ramsey, and Postma, 2006; Rapcsak, Ochipa, Anderson, and Poizner, 1995; Roland, Larsen, Lassen, and Skinhoj, 1980; Sakata, Shibutani, and Kawano, 1980). In line with this hypothesis, Yamakawa and colleagues' (2009) findings provided evidence in favor of the existence of a common neural substrate in the parietal cortex for both mental representations of social relationships and physical space.

Further supporting the view that physical and psychological spaces are inextricably linked, is the observation, now dating back over fifty years, that the distance between individuals varies as a function of their intimacy (see, e.g., Hall, 1964, 1969). One of the most interesting and fundamental pillars of *Proxemics* − the study of personal space (Argyle and Dean, 1965; Hall et al., 1968; Hayduk, 1983) indicates that people unconsciously organize the space around them in concentric areas, so that the areas closest to one's body are the privileged space of action(s) for the most intimate interactions, and, conversely, the areas most distant from the body are mostly associated with the space of action(s) for interactions with individuals with whom they share a low degree of intimacy. These concentric "virtual" zones around the individual's body may vary according to different factors, such as the culture or the gender of the individuals, but the general principle according to which a relationship exists between the degree of intimacy between two individuals and the physical distance that tends to settle during their interaction is a constant element independent of other factors (Hall, 1964).

These considerations on the direct relationship between physical and psychological distance led us to hypothesize that empathy toward others' pain could be modulated also on the basis of the physical distance between the observer and the individual subjected to a pain stimulation, just as happens for the social and affective distance (Avenanti et al., 2010; Sessa, et al., 2014; Singer et al., 2004, 2006; Xu et al., 2009).

In order to test this hypothesis, in Experiment 1 two groups of participants were administered a pain decision task (Sessa, et al., 2014; Sessa, et al., , 2014; Xu et al., 2009) in which faces (either upright or inverted) were presented in two different experimental conditions, i.e. pricked by a syringe (pain condition) or touched by a Q-tip (neutral condition), while participants' electroencephalogram (EEG) was recorded. Participants' task was to decide whether each face was painfully or neutrally stimulated. Importantly, the two groups of participants were presented with face stimuli of one of two possible sizes in order to manipulate the retinal size and therefore the perceived physical distance (see, e.g., Gogel, 1998), which approximately corresponded to close (6.56 feet, approximately 2 meters/close social distance) and far (9.84 feet, approximately 3 meters/far social distance) portions of the social distance (Hall, 1964). The choice to select these two specific perceived distances was based on the organization of the concentric "virtual" zones identified by the Proxemics. In particular, we decided to choose distances that were attributable to the "zone" ascribed to the "social distance" as identified by Hall (1963). This zone is located beyond the personal space that is reserved for more or less intimately known people, and is instead reserved for strangers, people one has just met and acquaintances. Since the faces that participants observed in this study were all of strangers, we considered it more appropriate from an ecological point of view that they were presented within the social distance zone. Furthermore, we have avoided presenting faces at a perceived distance corresponding to the personal space since it is known that, when this space is invaded, affective states that are in contrast with a possible empathic reaction may occur in the observer, such as anxiety, distress or anger (Hall, 1969). The social distance, on the other hand, permits interaction with others, but allows at the same time the individual to feel safe. It is important to add that this social distance zone can be in turn divided into two different portions or *phases* affecting the (potential) interaction with others, one corresponding to the close social space (within 7 feet or 2.1 meters) and one corresponding to the far social space (over 7 feet, and up to about 12 feet or 3.7 meters). Therefore, in line with this body of knowledge, we decided to use two sizes of face stimuli corresponding to perceived distances within the close social zone and within the distant social zone.

We adopted a minimalist experimental manipulation to induce different perceived distances of the face stimuli in order to keep stimulation as similar as possible to that usually employed in the standard pain decision task, and to limit the introduction of confounding elements, as for example other stimuli in the visual scene in addition to the empathy-related stimuli, possibly able to affect event-related potentials (ERPs) in unpredictable ways. On the other hand, if an object's size is known, as for faces, its retinal image can be used to judge its distance (see, e.g., Gogel, 1998).

The usually observed ERPs modulations observed in the pain decision task involve a shift towards more positive values for the pain condition than the neutral condition of a subset of ERPs components ranging from the P2 to the P3/LPP components recorded at both frontal and parietal electrode sites (Sessa and Meconi, 2015; Sheng et al., 2016; see also, e.g., Donchin, 1981; Donchin and Coles, 1988; Sessa, Luria, Verleger, and Dell'Acqua, 2007; Verleger, 1988). An ERP empathic reaction is defined by the difference between ERP(s) elicited in the pain and in the neutral conditions (Decety, Yang, and Cheng, 2010; Fan and Han, 2008a; Li and Han, 2010; Sessa, et al., 2014; Sheng and Han, 2012). In Experiment 1, we expected to observe a moderating effect on empathic ERP reactions as a function of the perceived physical distance of the faces, such that the group of participants exposed to faces perceived as more distant would have manifested a lower magnitude of these neural empathic reactions when compared to the group of participants exposed to faces perceived as closer. We hypothesized that inverted faces would not have induced an empathic reaction because of the disruption of the configural/holistic processing (Leder and Bruce, 2000) in either groups of participants. For this reason we expected reduced if null empathic reactions for inverted faces for both groups of participants. In this vein, we considered the inverted face condition that served as a control for other possible intervening factors in modulating ERPs. However, to our knowledge this is the first study investigating whether inverted faces painfully or neutrally stimulated may induce or not empathic reactions, therefore this aspect of the present study was purely exploratory.

We further designed a second experiment (Experiment 2) to test whether possible modulations of the neural empathic reactions in Experiment 1 could be ascribable to differences in the ability to identify faces of the two different sizes. In order to investigate this possibility, in Experiment 2, a new group of participants was engaged in a behavioural match-to-sample task involving the two-size upright face stimuli of Experiment 1.

## 2. Experiment 1

### 2.1. Method

#### 2.1.1. Participants

Before starting data collection, we established to enter into ERP analyses data from 15-20 participants for each of the two experimental groups because of existing literature in this field that suggests it is an appropriate sample (Fan and Han, 2008b; Sheng and Han, 2012). Analyses were conducted only after data collection was complete. Data were then collected from 40 volunteer healthy students (11 males) from the University of Padova. Data from 7 participants were excluded from the analyses due to excessive electrophysiological artifacts, of which 17 for one group and 16 for the other group. For this reason an additional participant was tested such that the two groups had the same number of participants. All participants reported normal or corrected-to-normal vision and normal audition and no history of neurological disorders. They were randomly assigned to the two different groups, as a function of the two different physical sizes of face stimuli. Each group included 17 participants (for far physical distance: 5 males; mean age: 23.8 years, SD =; 4.28, 4 left-handed; for close physical distance: 6 males, mean age: 23.2 years, SD =; 3.62, 4 left-handed). All participants signed a consent form according to the ethical principles approved by the University of Padova.

2.1.2. Stimuli

The stimuli were 12 digital photographs of White faces with a neutral facial expression from the Eberhardt Lab Face Database (Mind, Culture, & Society Laboratory at Stanford University, http://www.stanford.edu/group/mcslab/cgi-bin/wordpress/examine-the-research/). Each face was digitally manipulated in order to obtain stimuli for two different stimulation conditions, one in which faces received a painful stimulation (needle of a syringe penetration), and one in which faces received a neutral (Q-tip touch) stimulation (applied either to the left or to the right cheek).

All faces were presented in the upright and inverted orientation and in two different physical sizes, in order to manipulate retinal size and perceived physical distance, both beyond the intimate and personal distances, and roughly corresponding to the close and far portions of social distance (Hall, 1964). Face stimuli appearing to be in the far portion of social distance fit in 1.6° x 2.5° (width x height), whereas face stimuli appearing in the close portion of social distance fit in 2.5° x 3.3° (width x height). One group was exposed to faces appearing to be distant from participants 6.56 feet (approximately 2 meters; close social distance) and the other group was exposed to faces appearing to be distant from participants 9.84 feet (approximately 3 meters; far social distance). Stimuli were presented on a 17-in cathode ray tube monitor controlled by a computer running E-prime software.

### 2.1.3. Experimental design

We implemented a variant of the pain decision task. Each trial began with the presentation of a fixation cross at the centre of the screen (800–1600 ms, jittered in steps of 100 ms), followed by a face displayed for 400 ms. The sequence of events of each trial is depicted in Figure 1. Please note that the original face stimuli have been replaced in Figure 1 (a and b) with other face stimuli not belonging to the Eberhardt Lab Face Database according to the terms of use of the Database.

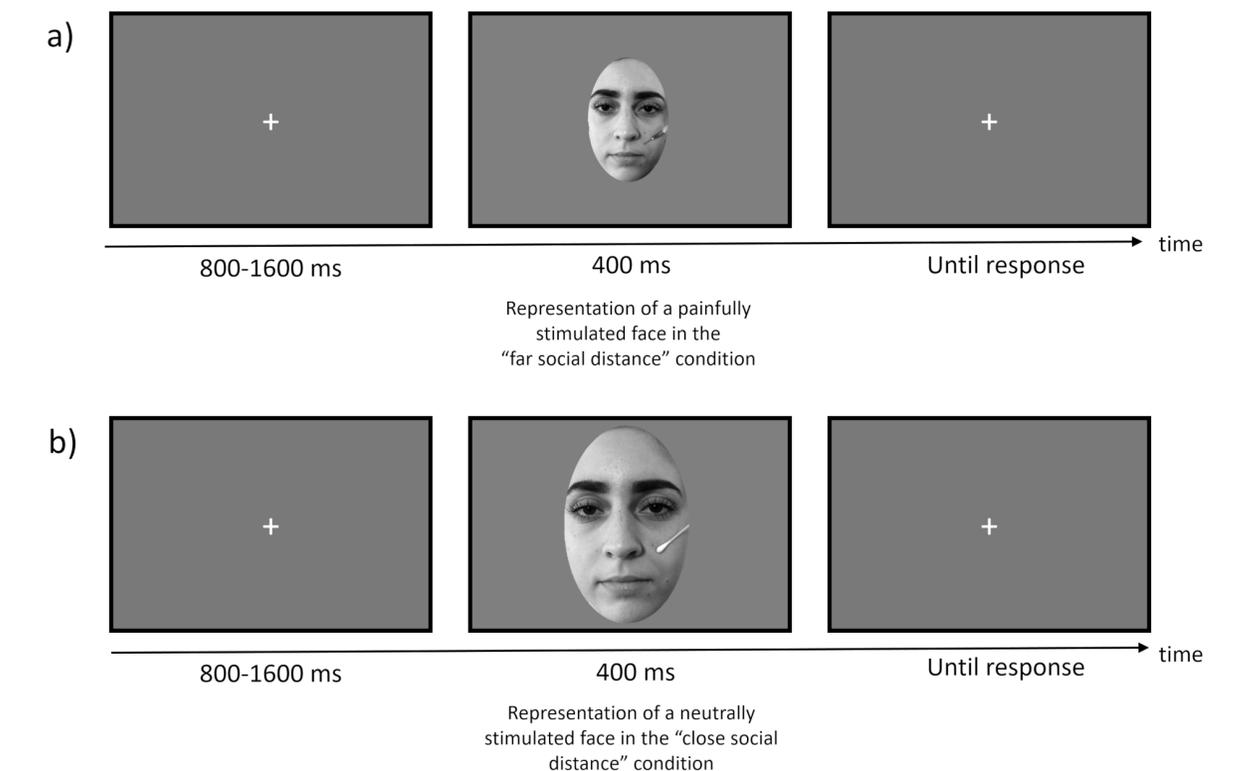

Participants were instructed to decide whether each face was painfully or neutrally stimulated by pressing one of two appropriately labelled keys of the computer keyboard as quickly and accurately as possible. Following a brief session of practice in order to familiarize with the task, participants performed 576 trials divided in 4 blocks (144 trials for each block including all the possible experimental combinations, intermixed within each block). Participants could manage a break session between a block and the next block of trials and decided when to continue by pressing the space bar. The experiment lasted for approximately 30-40 minutes. The entire experimental session, including the preparation of the participant for the EEG data collection, lasted about 60-75 minutes.

2.1.4. Electrophysiological recording and analyses

The EEG was recorded from 64 active electrodes distributed over the scalp in accordance with the international 10/20 system placed on an elastic Acti-Cap, referenced to the left earlobe. The EEG was re-referenced offline to the average of the left and right earlobes. Horizontal EOG (i.e., HEOG) was recorded bipolarly from two external electrodes positioned laterally to the left and right external canthi. Vertical EOG (i.e., VEOG) was recorded from Fp1 and one external electrode placed below the left eye. The electrode impedance was kept less than 10 KΩ because of the highly viscous electro-gel and the properties of active electrodes. Offline EEG processing and analyses were conducted using Brain Vision Analyzer software (Brain Products; www.brainproducts.com).

EEG, HEOG and VEOG signals were amplified (pass band 0.01–80 Hz) and digitized at a sampling rate of 250 Hz. The EEG was segmented into 1200 ms epochs starting 200 ms prior to the onset of the faces. The epochs were baseline-corrected based on the mean activity during the 200 ms pre-stimulus period, for each electrode site. Trials associated with incorrect responses or contaminated by large horizontal eye movements, eye blinks or other artifacts (exceeding ± 30 µV, ± 60 µV and ± 80 µV, respectively) were automatically discarded from analysis, which accounted for the exclusion of an average of 6% of trials. Separate average waveforms for each condition were then generated time-locked to the presentation of the face stimuli for each experimental condition. Statistical analyses of ERPs mean amplitudes focused on a time window ranging from 300 and 600 ms, corresponding to the P3 ERP component. Mean P3 amplitude values were measured at pooled electrode sites selected from fronto-central (Fz, F1, F2, F3, F4, F5, F6, FCz, FC1, FC2, FC3, FC4,

FC5, FC6) and centro-parietal (CPz, CP1, CP2, CP3, CP4, CP5, CP6, Pz, P1, P2, P3, P4, P5, P6) electrodes according to visual inspection and previous work (Fan and Han, 2008b; Meconi et al., 2015; Sessa and Meconi, 2015; Sessa, et al., 2014).

2.1.5. Behavioral and ERPs results

The significant threshold for all statistical analyses was set to .05. Exact *p* values and effect sizes (i.e., partial eta-squared, $\eta p^2$) are reported. Planned comparisons relevant to test the hypotheses of the present experiment are reported.

*Behavioral results*

Individual mean proportion of correct responses was submitted to a mixed analysis of variance (ANOVA), considering stimulation of face stimuli (painfully vs. neutrally stimulated) and orientation (upright vs. inverted) as within-subjects factors and physical distance (far social distance vs. close social distance) as a between-subjects factor. The main effect of neither face stimuli or orientation were significant (respectively: $F < 1$, $p = .970$, $\eta p^2 = .000$; $F(1,32) = 3.679$, $p = .064$, $\eta p^2 = .103$); the mean proportion of correct responses for face stimuli neutrally stimulated in the upright orientation was .984; $SD = .17$, and in the inverted orientation condition was .985; $SD = .14$; the mean proportion of correct responses for face stimuli painfully stimulated in the upright orientation was .985; $SD = .17$, and in the inverted orientation was .9.88; $SD = .14$). The interactions between face stimuli and physical distance and between orientation and physical distance were not significant: $F < 1$, $p = .986$, $\eta p^2 = .000$; $F < 1$, $p = .341$, $\eta p^2 = .028$, respectively.

Reaction times (RTs) exceeding each individual mean RT in a given condition ± 2.5 *SD* and RTs associated with incorrect responses were excluded from the RTs analysis. Individual mean proportion of correct responses and RTs were submitted to a mixed ANOVA, including face stimuli (painfully vs. neutrally stimulated) and orientation (upright vs. inverted) as within-subjects factors and physical distance (far social distance vs. close social distance) as a between-subjects factor. None of the effects were statistically significant ($F < 1$; min $p = 0.98$).

*ERPs*

Grand averages of the face-locked ERP waveforms elicited in the pain and neutral stimulation conditions separately for pooled fronto-central (FC) and centro-parietal (CP) electrode site and for close and far social distance are shown in Figure 2 (upright face stimuli) and Figure 3 (inverted face stimuli).

A mixed analyses of variance (ANOVA) of P3 amplitude values including stimulation of face stimuli (painfully vs. neutrally stimulated) and orientation (upright vs. inverted) as within-subjects factors and physical distance (far vs. close) as a between-subjects factor was carried out for each ERP electrodes pool.

The ANOVA revealed a significant main effect of orientation at FC pooled electrode sites, $F(1,32) = 18.610$, $p < .001$, $\eta p^2 = .368$, and at CP pooled electrode sites, $F(1,32) = 16.908$, $p = .001$, $\eta p^2 = .514$). The main effect of stimulation of face stimuli reached significance level only for CP pooled sites, $F(1,32) = 7.950$, $p = .012$, $\eta p^2 = .332$, (at FC pooled sites: $F < 1$). The interaction between these two factors did not reach significance level for neither of the two pooled electrode sites (FC pooled sites: $F(1, 32) = 1.735$, $p = .206$, $\eta p^2 = .98$; CP pooled sites $F < 1$). Notably, the interaction between stimulation of face stimuli and physical distance reached significance both at FC pooled electrode sites, $F(1,32) = 8.697$, $p = .001$, $\eta p^2 = .020$, and at CP pooled electrode sites, $F(1,32) = 4.589$, $p = .040$, $\eta p^2 = .125$. Planned comparisons revealed that for face stimuli perceived at a closer physical distance the painful condition elicited more positive P3 amplitude than the neutral condition (at FC pooled sites: $t = -3.044$, $p = .008$; $Mdiff = -1.050$ [-1.78, -3.18]; at CP pooled sites: $t = -2.626$, $p = .018$; $Mdiff = -.915$ [.-1,65, -.176]). This effect was manifest as a positive shift of the ERP activity for face stimuli painfully stimulated (at FC pooled sites .964 µV, $SD = 2.34$; at CP pooled sites 4.98 µV, $SD = 3.48$) relative to face stimuli neutrally stimulated (at FC pooled sites -.0862 µV, $SD = 2.25$; at CP pooled sites 5.05 µV, $SD = 3.09$). Importantly, this positive shift indexing an empathic reaction was not observed for face stimuli appearing at far physical distance (at FC pooled sites: $t = 1.056$, $p = .307$; $Mdiff = .408$ [- .411, 1.22]; at CP pooled sites: $t = .188$, $p = .853$, $Mdiff = .069$ [.712, .8516]. The interaction between orientation and physical distance and the triple interaction between stimulation of face stimuli, physical distance and orientation were not significant (at FC: both $Fs > 1$; at CP: $F(1,32) = 1.444$, $p = .238$, $\eta p^2 = .043$ and $F < 1$, respectively).

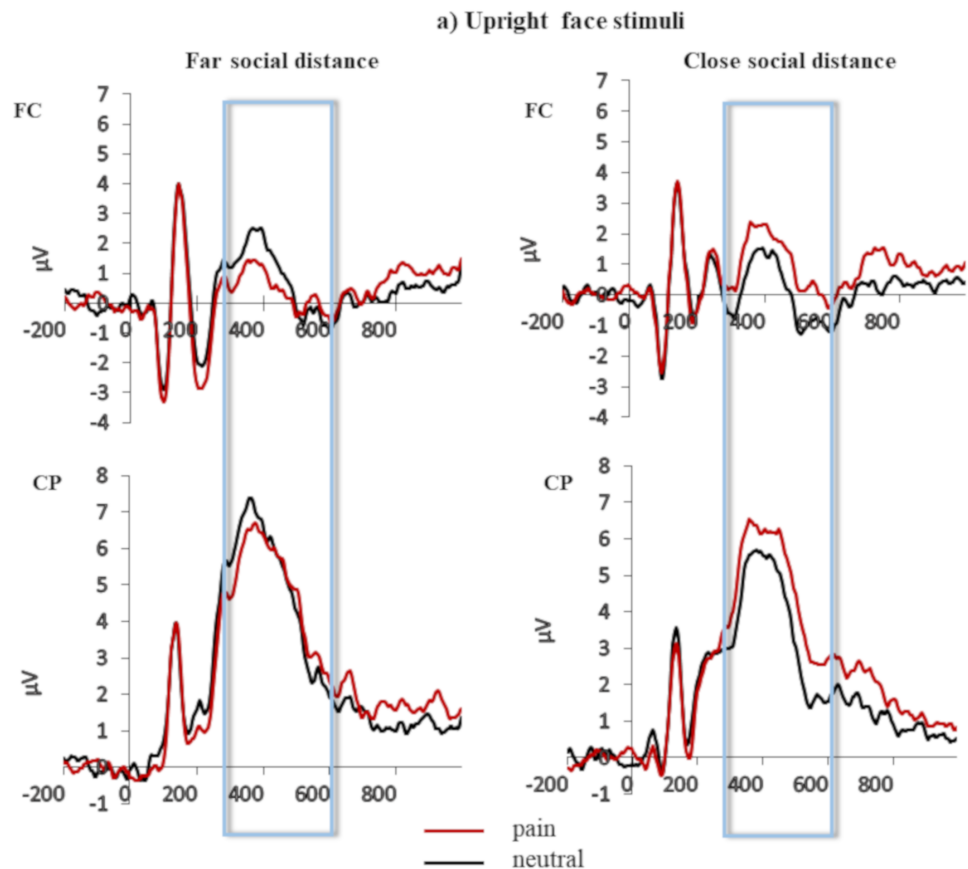

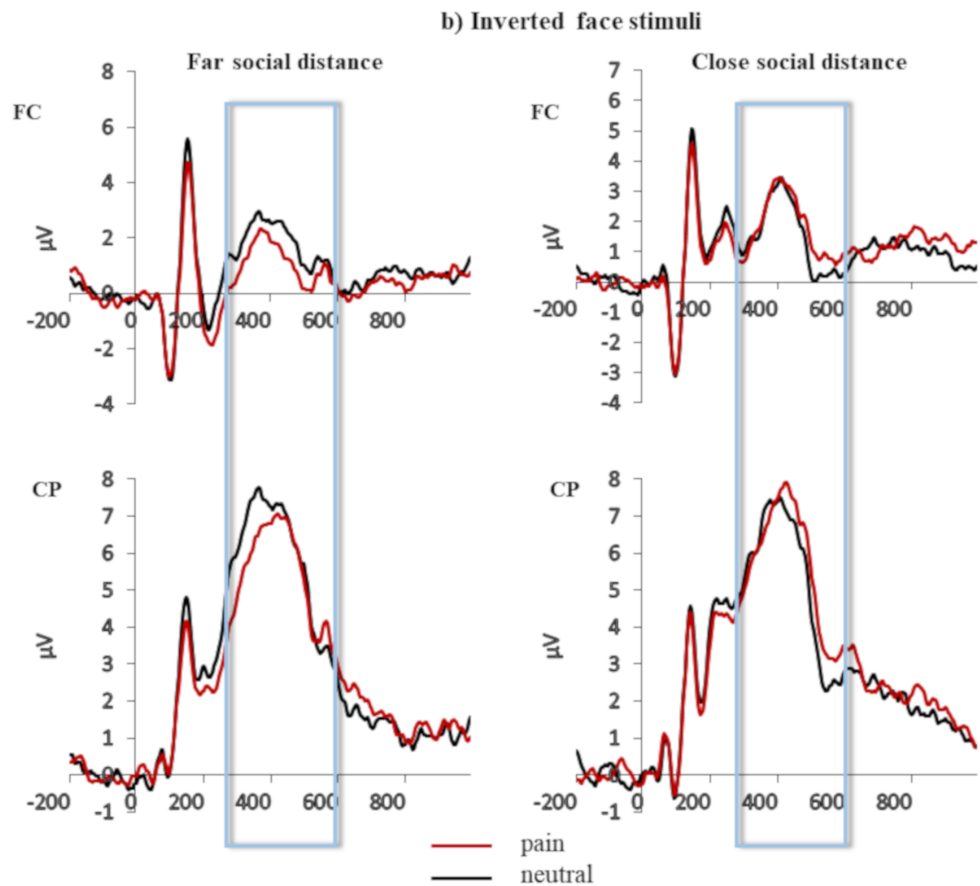

## 3. Experiment 2

Our experimental hypothesis on the modulating role of physical distance on empathy was corroborated, i.e. we observed greater empathic ERP reactions for the group of participants exposed to faces perceived as closer compared to the group of participants exposed to faces perceived as more distant, independently of faces orientation. As this first experiment left open the possibility that the differences observed between the two groups could depend on a different degree of discriminability of the faces perceived as closer and those perceived as more distant, we designed a second experiment (Experiment 2) to test whether the modulation of neural empathic reaction observed in Experiment 1 could be ascribable to differences in the ability to identify faces of the two different sizes. In order to investigate this possibility, in Experiment 2, a new group of participants was engaged in a behavioural match-to-sample task involving the two-size upright face stimuli of Experiment 1.

3.1. **Method**

3.1.1. Participants

Data were collected from 22 volunteer healthy students (3 males) from the University of Padova. All reported normal or corrected-to-normal vision and no history of neurological disorders. All 22 participants (3 males; mean age: 23.40 years, SD =1.79; 3 left-handed) were included in the final sample. All participants signed a consent form according to the ethical principles approved by the University of Padova. Analyses were conducted only after data collection was complete.

3.1.2. Stimuli

The stimuli were the same 12 digital photographs of White neutral faces from the Eberhardt Lab Face Database (Mind, Culture, & Society Laboratory at Stanford University, https://web.stanford.edu/group/mcslab/cgi-bin/wordpress/examine-the-research/) used in Experiment 1 (including the painful/neutral stimulation).

All faces were presented in the two different physical sizes used in Experiment 1. Stimuli were presented on a 17-in cathode ray tube monitor controlled by a computer running E-prime software.

3.1.3. Procedure

Experimental design

We implemented a variant of the discrimination task based on an XAB match-to-sample task used by Newell and Bülthoff (2002; see also Young et al., 1997). On each trial a face stimulus (stimulus X) was presented and then followed by two face stimuli (stimuli A and B) presented simultaneously, one on the left and one the right of the fixation.

Each trial began with a fixation cross presented for 500 ms. Then the first face stimulus (X) of one of the two possible sizes was shown for 750 ms in the center of the screen. The next pair of face stimuli (A and B), of the same size of the first face (stimulus X), remained on the screen until the participant pressed a response button. Each of the A and B face stimuli were displayed 3 cm to the left and to the right relative to the center of the screen.

Participants were instructed to respond as fast and as accurately as possible, indicating which face stimulus of the AB pair was identical to the preceding face stimulus X. Participants were instructed to press a key on the left (or on the right) of the keyboard to indicate that the face stimulus presented on the left (or on the right) was identical to the previously presented face stimulus (stimulus X). Following a brief session of practice in order to familiarize with the task, participants performed 528 trials, divided in 4 blocks (i.e., each block consisting of 132 trials). Faces of different sizes were presented in separate block of trials, whose order was counterbalanced between participants. Participants could manage a break session between the blocks trials and decided when to continue by pressing the space bar. The experiment lasted about 35 minutes.

Figure 4 shows two examples of trials, one (a) for the far social distance condition, and the other (b) for the close social distance condition. Original face stimuli have been replaced in the Figure 4 (a and b) with other face stimuli not belonging to the original database according to the terms of use of the Eberhardt Lab Face Database.

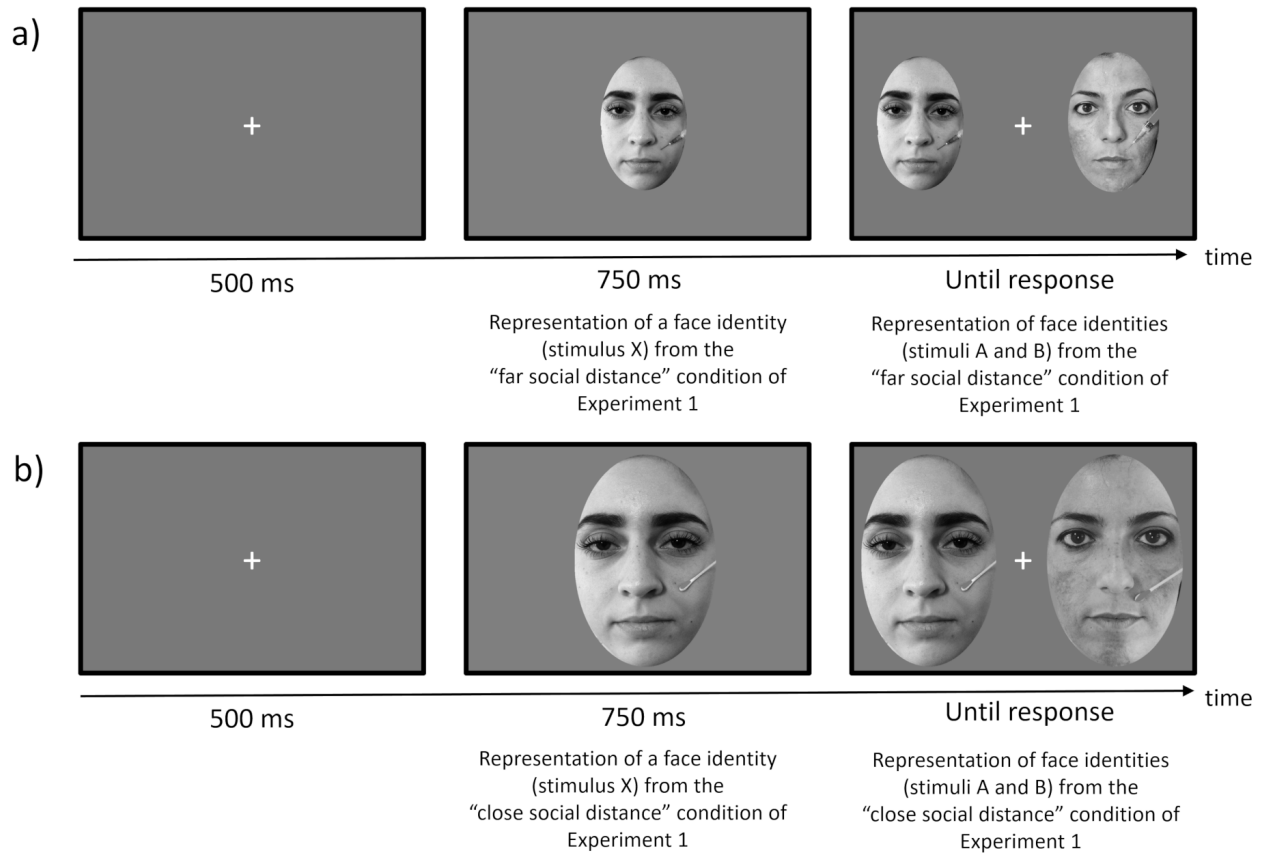

3.1.4. Statistical analysis

The significant threshold for all statistical analyses was set to .05. Exact *p* values and effect sizes (i.e., partial eta-squared, $\eta p^2$) are reported. Planned comparisons relevant to test the hypotheses of the present experiment are reported.

*Behavioral results*

Individual mean proportions of correct responses were submitted to a one-way analysis of variance (ANOVA), considering physical distance (far vs. close) as a within-subjects factor. The main effect of physical distance did not approach significance level: $F(1,21) = .236$, $p = .632$, $\eta p^2 = .11$ (see Figure 5).

RTs exceeding each individual mean RTs in a given condition ± 2.5 *SD* and RTs associated with incorrect responses were excluded from the RTs analysis. RTs were submitted to a one-way analysis of variance (ANOVA), considering physical distance (far vs. close) as a within-subjects factor. The effect of physical distance did not approach significant level: $F(1,21) = .648$, $p = .430$, $\eta p^2 = .030$.

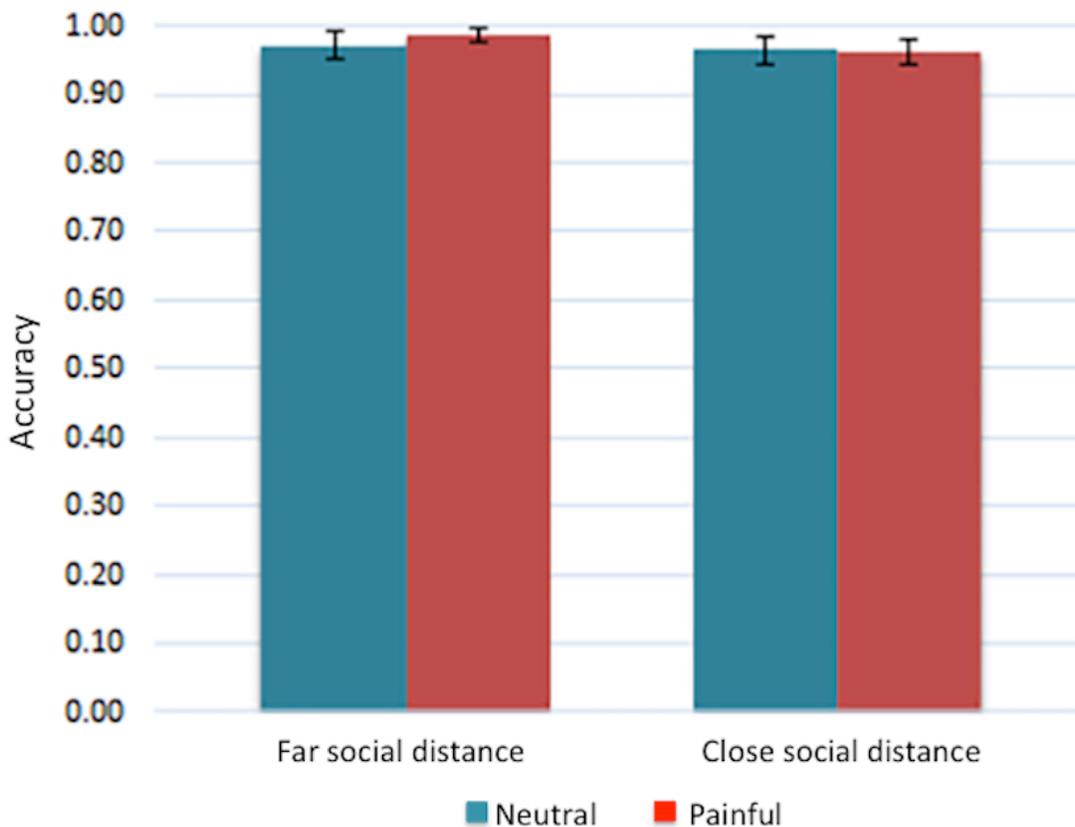

4. **Discussion**

A significant body of research has undoubtedly shown that the magnitude of an observer's empathic reaction depends on the social and affective bond existing with the individual experiencing an affective state in first-person (Avenanti et al., 2010; Contreras-Huerta et al., 2013, 2014; Jean

Decety and Svetlova, 2012; Lockwood, 2016; Rameson and Lieberman, 2009; Sessa et al., 2014; Singer et al., 2004; Xu et al., 2009). Based on robust experimental evidence suggesting the existence of an inextricable link between the processing of physical distance and that of psychological distance (Hall, 1964, 1969), the present study aimed at investigating whether physical distance, like the psychological distance, could be a modulator of the magnitude of the observer's empathic reaction for an individual in a state of physical pain. In order to test this hypothesis, we implemented a between-subjects experimental design (Experiment 1) in which we manipulated the perceived physical distance (close social distance: 6.56 feet, approximately 2 meters vs. far social distance: 9.84 feet, approximately 3 meters) of upright and inverted faces pricked by a syringe (i.e., pain condition) or touched by a Q-tip (i.e., neutral condition). We therefore expected to observe a reduced empathic reaction in the group of participants exposed to faces perceived as more distant when compared to the empathic reaction in the group of participants exposed to faces perceived as closer. Whether this reaction could be selectively observed for upright faces was an open question. In line with this hypothesis, the results indicated that in a time window between 300 and 600 ms following the presentation of the face stimuli, a clear ERP pattern previously linked with an empathic reaction (e.g., Meconi et al., 2015; Sessa et al., 2007; Sheng and Han, 2012) was observed at both fronto-central and centro-parietal regions in the group of participants exposed to the face stimuli perceived as closer, while this reaction was absent in the group of participants exposed to face stimuli perceived as more distant. This effect did not interacted with the orientation of the faces, suggesting that also inverted faces can elicit an empathic reaction. Importantly, no differences were observed in terms of accuracy in discriminating between the painful and the neutral stimulation conditions indicating that the differences in empathic reactions between the two groups of participants did not depend on differences in the ability to discriminate between the two stimulation categories (i.e., painful vs. neutral stimuli) in the two different sizes conditions, further suggesting that the observed differences in the empathic reaction depended indeed on the manipulation of perceived distance of face stimuli.

                                                                                     20

Experiment 1 did not allow us to clarify whether the modulation of the empathic reaction in the two groups depended on differences in discriminability of the faces of the two sizes. This possibility could be particularly relevant in light of the consolidated knowledge in the context of the social psychology of two possible putative cognitive operations that people use during the perception of others, i.e. individuation and categorization (see Brewer, 1988; Fiske and Neuberg, 1990). While individuation is that mechanism by which the other individual is perceived as a unique entity, the mechanism of categorization leads to others' perception based on their categorization as belonging to a specific social group. Notably, evidence in the context of empathy toward others' pain suggests that these mechanisms may be critical modulators of the empathic reaction, so that individuation favors an empathic reaction while categorization tends to be associated with its suppression (Sheng and Han, 2012). These considerations could therefore suggest that under conditions in which faces are more easily discriminable, an individuation mechanism can be favored and this in turn could promote an empathic reaction. We then implemented a second experiment (Experiment 2) that involved one further group of participants engaged in a behavioural match-to-sample task involving the same two-size upright face stimuli of Experiment 1 to test the hypothesis that the two categories of faces (perceived as closer and perceived as more distant) could be more or less easily identifiable. Results of Experiment 2 revealed that face stimuli of the two sizes could be equally identifiable both in terms of accuracy and reaction times, supporting the view that the critical factor triggering differential empathic reactions in the two groups of participants in Experiment 1 was not related to the likelihood of identifying the faces of the two sizes. We have to admit that this conclusion should be taken with caution because of the ceiling effect observed with regard to the accuracy level; however, we believe that the observation that also reaction times, that are characterized by a more meaningful variation, did not differ between the two sizes conditions provide additional support in favour of our interpretation. It is important to stress that this whole pattern of findings does not imply that an individuation mechanism may not be preferred for faces perceived as closer relative to those perceived as more distant, but rather that the implementation of this mechanism, rather than that of categorization, does not seem to be a direct consequence of the ease/difficulty of identifying faces.

We confess that we cannot rule out the possibility that the size of the faces *per se* (rather than distance perception) have produced those observed modulations in neural empathic reactions. Nevertheless, we believe this is unlikely since each face was neutrally or painfully stimulated by a tool that was proportional in size to the stimulated face, so the tool provided a contextual cue that participants could use to estimate distance. The findings that the two groups of participants were equally accurate and fast in deciding whether the faces were painfully or neutrally stimulated (Experiment 1) and in discriminating faces of the two sizes (Experiment 2) strongly support the idea that it was not the size per se the key modulator factor of the empathic reactions but rather the perceived distance of the faces. Moreover, as already discussed in the Introduction section, for stimuli whose size is known and familiar to an observer, their size and the retinal image size are sufficient indications to induce an estimate of physical distance (see, e.g. Gogel, 1998).

Although the perception of distance has been proved a fundamental modulator of interpersonal processes, including empathy for pain as demonstrated in the present work, the underlying mechanism is not well understood. At least two classes of theories – that are not mutually exclusive – could account for this modulatory effect, i.e. the *Construal Level Theory* (CLT; Trope, Liberman et al., 2007) and the *Embodied Cognition Theory* (see, e.g., (Caruana and Borghi, 2013; Dijkstra, Kaschak, and Zwaan, 2007; Gallese, 2005; Goldman and de Vignemont, 2009; Niedenthal, 2007). The first theoretical approach suggests that as the physical, temporal, social and psychological distance between an individual and an event, an object, or even a person or a group of people increases, not only the salience and perceived relevance diminish (e.g., Latané, Liu, Nowak, Bonevento, and Zheng, 1995; Latané, 1981; Williams and Bargh, 2008), but also mental representations of events, objects and other people profoundly change so that as the distance

 

increases, the degree of abstraction of mental representations also increases (e.g., Henderson, Wakslak, Fujita, and Rohrbach, 2011). Notably, Williams et al. (2008) observed that among all of these different types of distances, physical distance is a sort of ontogenetic precursor of all of other types, "the foundation for the later-developed concept of psychological distance" (Williams and Bargh, 2008). Interestingly, this idea dovetails nicely with the evidence provided by the fMRI study by Yamakawa and colleagues (2009) presented in a previous paragraph suggesting a common neural underpinning for both psychological and physical distance representations in the parietal cortex. Moreover, in line with both the CLT and the experimental evidence provided by the present study, the previous work by Williams and Bargh (2008) had shown, through the implementation of 4 experiments, that when people are exposed to cues of physical distance these can have a moderating effect on their emotional experience, for instance by modulating the degree of emotional attachment to family members or by reducing the level of emotional distress to the vision of violent media. These results converge with the finding that physical distance can therefore also play an important role in moderating an observer's empathic reaction toward others' pain.

    According to the theories of Embodied Cognition, most of the cognitive processes depend, reflect, or are influenced by the body control systems (e.g., Caruana and Borghi, 2013). Cognition would therefore be inextricably linked to the body and to its relation with the environment, and it would not be based on abstract and amodal representations. At least three different interpretations of how embodiment might influence cognition have been proposed (see Goldman and de Vignemont, 2009). According to a first interpretation, the body anatomy itself would play a role in cognition, precisely because of the anatomical characteristics of the different body parts. A second interpretation considers how the actions produced by the body can have a deep influence on cognitive processes (e.g., Dijkstra et al., 2007; Niedenthal, 2007); for example, posture and facial expressions could influence the way people remember, discriminate between different categories of stimuli, and

could even influence their emotional state. A third interpretation of embodiment, proposed and termed by Gallese (2005) *Embodied Simulation*, refers to the role that mental representations involving the body can have on cognition. This last interpretation of embodiment is strongly associated with the construct of empathy, and several authors, more or less explicitly, have suggested that embodied simulation/mirroring mechanisms are at the basis of the most automatic component of empathy (Csibra, 2008; Gallese, Migone, and Eagle, 2006; Gallese, 2003, 2008; Gallese and Goldman, 1998; Hickok, 2009; Lamm and Singer, 2010; Singer and Lamm, 2009; Uithol, van Rooij, Bekkering, and Haselager, 2011; but see also Lamm and Majdandžić, 2015a). Caggiano et al., (2009) have shown the existence of a subpopulation of mirror neurons in the premotor cortex of rhesus monkeys whose activity is modulated on the basis of the spatial position in which the observed action occurs; in particular, half of these neurons are activated preferentially for the monkey's peripersonal space while the other half is more responsive for the extrapersonal space. The authors interpreted these fascinating results by suggesting that mirror neurons (and likely, more generally, mirror mechanisms) not only constitute the neural substrate of the "understanding of *what others are doing*, but also may contribute toward selecting *how I might interact with them*" (Caggiano et al., 2009). This result could suggest that the neurally instantiated *we-centric space* (Gallese, 2003) underlying the embodied simulation − conceived as the mechanism that mediates our ability to share the meaning of actions, emotions, emotional states with others − might be sensitive to the physical distance that separates the observer and the other individual and to the space of potential interaction between the two, the so-called *interaction space*, that is the shared reaching space of the two individuals (Nguyen and Wachsmuth, 2011). These findings and observations could allow to predict that even the empathic reactions of an observer could be influenced by the distance that separates her/him from the individual experiencing a particular affective state and that these reactions might be different when the two individuals are within the space of potential interaction or not. We acknowledge that at the moment this second interpretation regarding the mechanism underlying the effect of physical distance in the modulation of empathy is certainly speculative (although intriguing) and will require further research.

Finally, we want to mention that our findings are in line with the evidence reported by Yang and colleagues (2014) that the efficiency of faces recognition, for both upright and inverted faces, varies as a function of faces size. The authors manipulated faces size between 1° and 10° of visual angle and demonstrated that only faces larger than 6° of visual angle are associated with the recruitment of specialized face processes. Additionally, while for faces smaller than 6° of visual angle (corresponding to a perceived distance of 2 m), only a quantitative difference between upright and inverted faces was observed in the recruitment of these processes, for faces larger than 6° of visual angle the difference was qualitative. The authors note that the distance of 2 m corresponds to the typical interpersonal distance in the context of conversations and social interactions. In brief, their findings support the notion that faces can be processed either through generic recognition processes or involve specialized face-sensitive processes depending on their perceived distance. Interestingly, the perceived distance of the larger faces used in our study corresponds to the upper limit indicated by Yang and colleagues. Finally, the evidence reported by Yang and colleagues also dovetails nicely with the mechanisms underlying CLT and embodied simulation as discussed in the previous paragraphs.

Lastly, we would like to discuss a few possible limitations of the present study. We implemented a between-subjects design (Experiment 1) that has less statistical power than within-subjects designs; between-subjects designs may also have the disadvantage that results may in part depend on inter-individual differences that may then characterize the two groups of participants differently. Nevertheless, the within-subject designs have few disadvantages that in the present experimental context we considered to be more alarming. In particular, the main weakness of within-subject designs is that they can be associated with carryover effects. These include effects of practice and fatigue, but in particular we wanted to avoid the "context effect", namely the effect for which stimuli that are perceived/evaluated in an experimental condition can alter how they are perceived/evaluated in a subsequent experimental condition. Obviously this possible effect could have greatly reduced if not eliminated the effects related to the manipulation of the perceived distance. Furthermore, a within-subjects manipulation of the variable relative to the size of the faces would have required doubling the number of trials for each participant in order to guarantee a sufficient signal-to-noise ratio, inevitably producing fatigue with potential electrophysiological effects. We have however tried to make the two groups homogeneous by age and gender, two of the variables that could have an impact on participants' empathy (for the age variable see, e.g., Phillips, MacLean, and Allen, 2002; Schieman, 2000) but see also Grühn, Rebucal, Diehl, and Labouvie-Vief, 2008; for the gender variable see, e .g., Cohn, 1991; Brown, 2001; Eisenberg and Lennon, 1983; Feingold, 1994; O'Brien, Konrath, Grühn, and Hagen, 2013; Thompson and Voyer, 2014; but see also Lamm, Batson, and Decety, 2007 for contrasting findings).

Furthermore, in the present investigation each experimental group consisted mostly of female participants. Previous studies, as briefly mentioned above, suggested that women's empathy might be greater than that of men and therefore the present results might not be straightaway generalizable to the entire population. Nevertheless, we note that precisely because of the greater empathic abilities found in women in previous studies, the ample reduction of the neural empathic response observed for the faces perceived as more distant is even more reliable.

Finally, we would like to briefly discuss about the statistically null triple interaction between stimulation of face stimuli, physical distance and orientation. A significant triple interaction would probably have further corroborated our conclusions that the perceived distance of someone in conditions of physical suffering is an important modulator of the observer's neural empathic response. However, it is important to underline, as already briefly mentioned in the Introduction, that

 

there are no previous studies, at least to our knowledge, that have directly tested empathic responses for inverted faces. In this vein, our results suggest that inverted faces may still be associated with a neural empathic response although we cannot rule out the possibility that this null result was due to an insufficient statistical power. On the other hand, in our opinion, the most striking and interesting finding of the present study is that linked to the interaction between stimulation of face stimuli, and physical distance, which support the conclusion that the perceived distance is an important factor able to modulate observer's empathy. To note, physical distance did not interact with orientation, narrowing the impact of physical distance on how the brain process painful vs. neutral stimulations (but not other characteristics of the faces such as their orientation).

In conclusion, in the present investigation we provided evidence that also the physical distance between an observer and another individual in a particular affective state − such that induced by physical pain − is a decisive factor for the modulation of an empathic reaction in the observer. This evidence provides an important insight into the framework of knowledge on factors capable of shaping empathy, and it is certainly important also in relation to the evidence suggesting a strong link between representations, also in neural terms, of physical and psychological distance. Although it is obvious that in everyday life situations it is not possible to establish in advance the physical distance between an observer and someone subjected to physical pain (given the unpredictability of such situations), the evidence on the importance of physical distance in modulating an empathic reaction could be fundamental for psychotherapy, clinical and medical contexts, in which psychotherapists, doctors and health professionals could use this knowledge to favor or not, as appropriate, an empathic reaction in themselves and in their patients.

*E Scienze Umane*, *40*(3), 543–580.

Gogel, W. C. (1998a). An analysis of perceptions from changes in optical size. *Perception & Psychophysics*, *60*(5), 805–20. doi.org/10.3758/BF03206064

Gogel, W. C. (1998b). An analysis of perceptions from changes in optical size. *Perception & Psychophysics*, *60*(5), 805–20.

Goldman, A. I., & de Vignemont, F. (2009). Is social cognition embodied? *Trends in Cognitive Sciences*, *13*(4), 154–159. doi.org/10.1016/j.tics.2009.01.007

Grühn, D., Rebucal, K., Diehl, M., & Labouvie-Vief, G. (2008). Empathy Across the Adult Lifespan: Longitudinal and Experience- Sampling Findings. *Emotion*, *8*(6), 753–765. doi.org/10.1037/a0014123.Empathy

Hall, E. T. (1963). System Notation Proxemic. *American Anthropologist,* 1003–1025.

Hall, E. T. (1969). *The Silent Language*. (Doubleday&Company, Ed.). Garden City, New York: Anchor Books.

Hall, E. T., Birdwhistell, R. L., Bock, B., Bohannan, P., Richard, A., Durbin, M., … Hall, E. T. (1968). Proxemics [and Comments and Replies]. *Current Anthropology*, *9*(23), 83–108.

Harris, L. T., & Fiske, S. T. (2006). Dehumanizing the Lowest of the low. *Psychological Science*, *17*(10), 847–853. doi.org/10.1111/j.1467-9280.2006.01793.x

Hayduk, L. A. (1983). Personal space: Where we now stand. *Psychological Bulletin*, *94*(2), 293–335. doi.org/10.1037/0033-2909.94.2.293

Hein, G., Silani, G., Preuschoff, K., Batson, C. D., & Singer, T. (2010). Neural responses to ingroup and outgroup members' suffering predict individual differences in costly helping. *Neuron*,
This is a provisional file, not the final typeset article  32

**Author contributions**

P.S. developed the study concept. All authors contributed to the study design. A.S.L., F.M. and I.R. performed testing and data collection. P.S. and A.S.L. performed the data analysis and interpreted the data. P.S. and A.S.L. drafted the manuscript. All authors approved the final version of the manuscript for submission.

**Competing financial interests**

The authors declare no competing financial interests.

**Figure Captions**

Figure 1. Timeline of each trial for Experiment 1 (pain decision task): a) example of a trial for the far social distance condition with a neutrally stimulated face; b) example of a trial for the close social distance condition with a painfully stimulated face. Original face stimuli have been replaced in the Figure 1 (a and b) with actors according to the terms of use of the Eberhardt Lab Face Database.

Figure 2. Grand averages of the face-locked ERP waveforms for the upright face stimuli elicited in the pain and neutral stimulation conditions separately for pooled fronto-central (FC) and centro-parietal (CP) electrode site and for close and far social distance.

Figure 3. Grand averages of the face-locked ERP waveforms for the inverted face stimuli elicited in the pain and neutral stimulation conditions separately for pooled fronto-central (FC) and centro-parietal (CP) electrode site and for close and far social distance.

Figure 4. Timeline of each trial for Experiment 2 (match-to-sample task): a) example of a trial with faces used for the "far social distance" condition of Experiment 1; b) example of a trial with faces used for the "close social distance" condition of Experiment 1. Original face stimuli have been replaced in the Figure 4 (a and b) with actors according to the terms of use of the Eberhardt Lab Face Database.

Figure 5. Bar chart displaying mean rating scores for each condition for Experiment 2. Error bars represent standard errors.